\documentstyle[12pt]{article}
\newcommand{\ket}{\rangle}
\newcommand{\R}{\mbox{\boldmath $ R $}}
\newcommand{\Z}{\mbox{\boldmath $ Z $}}

\makeatletter
\@addtoreset{equation}{section}
\makeatother

\begin{document}
%
\baselineskip 6mm
\begin{flushright}
KU-AMP 95015 \\
hep-th/9601079 \\
January 1996 \\
revised in March 1996
\end{flushright}
\vspace*{10mm}
\baselineskip 8mm
\begin{center}
{\Large %
Algebra and Twisted Algebra in Toroidal Target Space}\footnote{%
Submitted to Mod. Phys. Lett. A}
\vspace{12mm}
\\
{\large \sc Shogo Tanimura}\footnote{%
e-mail address : tanimura@kuamp.kyoto-u.ac.jp}
\\
{\it 
Department of Applied Mathematics and Physics, \\
Faculty of Engineering, Kyoto University, \\
Kyoto 606-01, Japan
}
\vspace{16mm}
\\
{\bf Abstract}
\\
\begin{minipage}[t]{120mm}
\baselineskip 6mm
Target space duality is reconsidered from the viewpoint of
quantization in a space with nontrivial topology.
An algebra of operators for the toroidal bosonic string is defined
and its representations are constructed.
It is shown that there exist an infinite number of inequivalent quantizations,
which are parametrized by two parameters $ 0 \le s, t < 1 $.
The spectrum exhibits the duality only when $ s = t $ or $ -t $ (mod 1).
A deformation of the algebra by a central extension is also introduced.
It leads to a kind of twisted relation
between the zero mode quantum number and the topological winding number.
\end{minipage}
\end{center}
%
\newpage
\baselineskip 7mm
\section{Introduction}
Recently target space duality (T-duality) in string theory 
has attracted a lot of attention
and is being investigated
enthusiastically~\cite{Kikkawa,Sakai,Giveon,%
EAlvarez,OAlvarez1,OAlvarez2,Klimcik}.
Target space duality is an equivalence of two-dimensional sigma models
which are defined in target spaces 
with different geometry and sometimes different topology.
The simplest example of target space duality
is found in toroidal bosonic string model
which has a target space $ S^1 $ of radius $ R $.
Target space duality implies equivalence of two toroidal models;
although one has a radius $ R $
and the other one has a radius $ 1/( 2 \pi R) $,
their spectra coincide.

On the other hand, several authors~\cite{Mackey,Landsman,LL,OK,Tsutsui}
have investigated
the relation between geometry and quantum theory
in a context different from string theory.
They tried to construct 
quantum mechanics on a manifold with nontrivial topology by various methods.
They extended the canonical formalism of quantization
to include topological effects manifestly.
Then they noticed that 
there are an infinite number of inequivalent quantizations 
on a space with nontrivial topology.
Thus, although a beautiful understanding of 
target space duality is given~\cite{OAlvarez1,OAlvarez2}
in classical theory,
a richer structure in quantum theory of target space duality is expected,
which may have been missed before.
Target space duality is also investigated
from the path integral point of view~\cite{Buscher,Tyurin}.

The aim of this letter is 
to explore the quantum aspect of target space duality
from the viewpoint of quantization on a topologically nontrivial space.
We reconsider the simplest model, the toroidal bosonic string.
We define the algebra of quantum operators to describe the model
while keeping its topological nature manifest.
In this definition we will show the possibility of modifying the algebra;
this modified algebra may be called an algebra with a central extension 
or a twisted algebra.
Then representations of the algebra are constructed
and it is shown that there are also an infinite number of
inequivalent quantizations,
which are parametrized
by two parameters $ s $ and $ t $ $ ( 0 \le s, t < 1) $.
It is also shown that there is a target space duality
only when $ s = t $ or $ -t $ (mod 1).
Hence the concept of target space duality
may become subtler and richer in quantum theory than in classical theory.
This letter is an extension of a previous work~\cite{sigma},
in which only one central extension was considered.
Here we take into consideration all possible central extensions.
%
\section{Quantum algebra for the toroidal boson}
In classical theory the toroidal boson model is described 
by a dynamical variable which is a map
$ X : S^1 \to T^n, \: \sigma \mapsto X^A ( \sigma ) $.
For simplicity, we concentrate on the circular boson; 
$ X : S^1 \to S^1, \: \sigma \mapsto X( \sigma ) $.
Moreover we assume that the radius of the target space is normalized
to be of unit length.
Eventually we will replace it with a parameter $ R $
and we will consider the target space duality in this context.

To be a closed string, $ X( \sigma ) $ should satisfy the boundary condition
$ X( \sigma + 2 \pi ) = X( \sigma ) + 2 \pi n , \: ( n \in \Z ) $.
However the variable $ X( \sigma ) $ is not suitable 
to describe quantum theory since it is multi-valued.
To make a concise description of the quantum theory,
it is desirable to find a single-valued operator instead of $ X( \sigma ) $.
Thus we introduce
$ \hat{V} ( \sigma ) = e^{ i \hat{X} ( \sigma ) } $,
which would be a unitary operator.
Rigorously speaking,
to make sense of the exponential of the local operator $ \hat{X}( \sigma ) $,
we must treat the divergence accompanying it.
However it is treated by the rather standard normal ordering procedure,
so we will skip an argument to justify it.

$ \hat{X} ( \sigma ) $ is assumed to satisfy the canonical commutation relation
\begin{equation}
	[ \, \hat{X}( \sigma ), \, \hat{P}( \sigma' ) \, ]
	=
	i \delta ( \sigma - \sigma' ),
	\label{2.1}
\end{equation}
where the delta function is defined on $ S^1 $.
The geometric meaning of the above algebra will become clear
if we introduce a unitary operator
\begin{equation}
	\hat{U}_f 
	= 
	\exp
		\left[
			- i \int_0^{2 \pi} f( \sigma ) \hat{P} ( \sigma ) d \sigma 
		\right]
	\label{2.2}
\end{equation}
for an arbitrary function $ f : \R \to \R $ satisfying
$ f( \sigma + 2 \pi ) - f( \sigma ) = 2 \pi m $ $ ( m \in \Z ) $.
Then it is easily seen that
\begin{eqnarray}
	&&
	\hat{U}_f^\dagger \hat{X}( \sigma ) \hat{U}_f
	=
	\hat{X}( \sigma ) + f( \sigma ),
	\label{2.3}
	\\
	&&
	\hat{U}_f^\dagger \hat{V}( \sigma ) \hat{U}_f
	=
	e^{ i f( \sigma ) } \hat{V}( \sigma ).
	\label{2.4}
\end{eqnarray}
 From these relations we may call $ \hat{U}_f $
the deformation operator of the string configuration.
$ \hat{U}_f $ also generates a topology change of the string
as a special case.
Let us define a winding operator
\begin{equation}
	\hat{W}
	= 
	\exp
		\left[
			- i \int_0^{2 \pi} \sigma \hat{P} ( \sigma ) d \sigma 
		\right].
	\label{2.5}
\end{equation}
Then it satisfies 
\begin{eqnarray}
	&&
	\hat{W}^\dagger \hat{X}( \sigma ) \hat{W}
	=
	\hat{X}( \sigma ) + \sigma,
	\label{2.6}
	\\
	&&
	\hat{W}^\dagger \hat{V}( \sigma ) \hat{W}
	=
	e^{ i \sigma } \hat{V}( \sigma )
	\label{2.7}
\end{eqnarray}
as the name of the winding operator indicates.
Additionally, we may define a translation operator
\begin{equation}
	\hat{T}_\lambda
	= 
	\exp
		\left[
			- i \lambda \int_0^{2 \pi} \hat{P} ( \sigma ) d \sigma 
		\right],
	\label{2.8}
\end{equation}
which automatically satisfies the following;
\begin{eqnarray}
	&&
	\hat{T}_\lambda^\dagger \hat{X}( \sigma ) \hat{T}_\lambda
	=
	\hat{X}( \sigma ) + \lambda,
	\label{2.9}
	\\
	&&
	\hat{T}_\lambda^\dagger \hat{V}( \sigma ) \hat{T}_\lambda
	=
	e^{ i \lambda } \hat{V}( \sigma ).
	\label{2.10}
\end{eqnarray}

Using these operators $ \hat{V}( \sigma ) $ and $ \hat{U}_f $,
instead of $ \hat{X}( \sigma ) $ and $ \hat{P}( \sigma ) $,
the fundamental algebra of the toroidal boson is written as
\begin{eqnarray}
	&&
	[ \, \hat{V}( \sigma ), \hat{V}( \sigma' ) \, ]
	= 
	0,
	\label{2.11}
	\\
	&&
	\hat{U}_f^\dagger \hat{V}( \sigma ) \hat{U}_f
	= 
	e^{ i f( \sigma ) } \hat{V}( \sigma ),
	\label{2.12}
	\\
	&&
	\hat{U}_f \hat{U}_g  
	= 
	e^{ - i c ( f, g ) } \hat{U}_{ f+g },
	\label{2.13}
\end{eqnarray}
where the phase factor $ e^{ - i c ( f, g ) } $ is inserted for generality.
This kind of generalization is often called a central extension.
Such a generalization is possible because the deformation operator
$ \hat{U}_f $ acts on $ \hat{V}( \sigma ) $
by the adjoint action in (\ref{2.12}).
Hence such an extra phase factor does not spoil the associative action
of deformations on the string configuration. Namely, we can deduce that
\begin{equation}
	\hat{U}_g^\dagger 
	\hat{U}_f^\dagger 
	\hat{V}( \sigma ) 
	\hat{U}_f
	\hat{U}_g
	= 
	\hat{U}_{ f+g }^\dagger 
	\hat{V}( \sigma ) 
	\hat{U}_{ f+g }.
	\label{2.14}
\end{equation}
However, since $ \hat{U}_f $ should satisfy the associativity
$ ( \hat{U}_{f_1}   \hat{U}_{f_2} ) \hat{U}_{f_3}     
=   \hat{U}_{f_1} ( \hat{U}_{f_2}   \hat{U}_{f_3} ) $,
the phase $ c ( f, g ) $ must satisfy a consistency condition
\begin{equation}
	  c( f_1 , f_2 )
	+ c( f_1 + f_2 , f_3 )
	= c( f_1 , f_2 + f_3 )
	+       c( f_2 , f_3 )
	\;\;\;
	( \mbox{mod} \: 2 \pi ).
	\label{2.15}
\end{equation}
Such a function $ c $ is called a 2-cocycle.
Possible 2-cocycles are already classified by Segal~\cite{Segal,Pressley}.
To describe them 
we decompose the function $ f_i ( \sigma ) $ $ ( i = 1, 2 ) $
into three parts as
\begin{equation}
	f_i ( \sigma ) = m_i \sigma + \lambda_i + \tilde{f}_i ( \sigma ),
	\label{2.16}
\end{equation}
where we define each term by $ 2 \pi m_i = f_i( 2 \pi ) - f_i( 0 ) $,
$ 2 \pi \lambda_i = \int_0^{ 2 \pi } ( f_i(\sigma) - m_i \sigma ) d \sigma $
and $ \tilde{f}_i( \sigma ) = f_i( \sigma ) - m_i \sigma - \lambda_i $,
respectively.
Then the 2-cocycles of Segal are defined by
\begin{equation}
	c_d ( f_1 , f_2 ) 
	= 
	d
	\left\{
		  m_1 \lambda_2 
		- m_2 \lambda_1 
		+ 
		\frac{1}{4 \pi} \int_0^{2 \pi} 
        \biggl(
        		  \frac{d \tilde{f}_1}{d \sigma} \tilde{f}_2 
				- \frac{d \tilde{f}_2}{d \sigma} \tilde{f}_1 
        \biggr)
        d \sigma
    \right\}.
    \label{2.17}
\end{equation}
They are specified by an integer $ d $, 
which is called the rank of the central extension.
%
\section{Representations with $ d = 0 $}
To complete the quantization of the toroidal boson
we will now construct representations of the fundamental algebra
(\ref{2.11})--(\ref{2.13}).
We factorize $ \hat{V}( \sigma ) $ and $ \hat{U}_f $ for convenience as
\begin{eqnarray}
	&&
	\hat{V} ( \sigma ) 
	= e^{ i \hat{N} \sigma } \hat{V} e^{ i \hat{x} ( \sigma ) },
	\label{3.1}
	\\
	&&
	\hat{U}_f =  \hat{W}^m e^{ -i \lambda \hat{P} } 
	\exp 
	\left[ 
		-i \int_0^{2 \pi} 
		\tilde{f}( \sigma ) \hat{\pi}( \sigma ) d \sigma 
	\right],
	\label{3.2}
\end{eqnarray}
where $ \hat{N} $, $ \hat{x} $, $ \hat{P} $ and $ \hat{\pi} $ are hermitian;
$ \hat{V} $ and $ \hat{W} $ are unitary operators.
Their geometric meaning is the following:
$ \hat{N} $ denotes the winding number;
$ \hat{V} 
= \exp [ i / ( 2 \pi ) \int_0^{2 \pi} 
( \hat{X}( \sigma ) - \hat{N} \sigma ) d \sigma ] $ represents the zero mode
which is a collective coordinate on the target space $ S^1 $;
$ \hat{x}( \sigma ) $ denotes the oscillatory degrees of freedom
satisfying $ \hat{x}( 2 \pi ) = \hat{x}( 0 ) $ and 
$ \int_0^{2\pi} \hat{x}( \sigma ) d \sigma = 0 $.
$ \hat{W} $ is the winding operator already mentioned;
$ \hat{P} = \int_0^{2 \pi} \hat{P} ( \sigma ) d \sigma $
is the zero mode momentum;
$ \hat{\pi}( \sigma ) $ is the canonical momentum conjugate to
the oscillator $ \hat{x}( \sigma ) $.

According to the above decomposition of operators,
the fundamental algebra (\ref{2.11})--(\ref{2.13}) is rewritten as
\begin{eqnarray}
	&&
    [ \, \hat{N} , \hat{W} \, ] = \hat{W},
    \label{3.4}
    \\
    &&
    [ \, \hat{P} , \hat{V} \, ] = \hat{V},
    \label{3.5}
    \\
    &&
    [ \, \hat{x} ( \sigma ) , \hat{\pi} ( \sigma' ) \, ]
    =
    i \Bigl( \delta( \sigma - \sigma' ) - \frac{1}{2 \pi} \Bigr)
    \label{3.6}
\end{eqnarray}
and all the other commutators vanish when $ d = 0 $.
The case of nontrivial central extension $ d \ne 0 $ will be considered 
in the next section.

Now representations of the algebra are easily constructed. 
 First, (\ref{3.4}) is represented by
\begin{eqnarray}
	&&
    \hat{N} \, | n + s \ket 
    = 
    ( n + s ) \, | n + s \ket,
    \label{3.7}
    \\
    &&
    \hat{W} \, | n + s \ket = | n + 1 + s \ket,
    \label{3.8}
\end{eqnarray}
where $ n $ is an integer and $ s $ is an undetermined real number.
The Hilbert space spanned by $ \{ | n + s \ket | \, n \in \Z \} $ 
is denoted by $ {\cal H}_{s} $.
For each value of $ s $ $ ( 0 \le s < 1 ) $, the corresponding
$ {\cal H}_{s} $ provides an inequivalent representation.
The appearance of inequivalent representations is a phenomenon
observed for quantization in a topologically nontrivial
space~\cite{Landsman,OK,S1}.

Second, (\ref{3.5}) is actually isomorphic to (\ref{3.4}),
so its representation is constructed in the same way;
\begin{eqnarray}
	&&
    \hat{P} \, | p + t \ket 
    = 
    ( p + t ) \, | p + t \ket,
    \label{3.9}
    \\
    &&
    \hat{V} \, | p + t \ket = | p + 1 + t \ket,
    \label{3.10}
\end{eqnarray}
where $ p $ is an integer and $ t $ is a real parameter.
Here the Hilbert space is denoted by $ {\cal H}_t $.
The zero mode momentum is discrete, as expected for a toroidal target space,
but it is shifted by a fraction $ t $ $ ( 0 \le t < 1 ) $.
The inequivalent representations are again characterized by $ t $.

Third, the canonical commutation relation (\ref{3.6})
is represented by the standard Fock representation;
we make a Fourier expansion
\begin{eqnarray}
	&&
    \hat{x} (\sigma)
    = 
    \frac{1}{2 \pi} \, \sum_{ k \ne 0 } \, 
    \sqrt{ \frac{ \pi }{ | k | } } \,
    ( \hat{a}_k         \, e^{   i k \sigma }
    + \hat{a}_k^\dagger \, e^{ - i k \sigma } ),
    \label{3.11}
    \\
    &&
    \hat{\pi} (\sigma)
    = 
    \frac{i}{2 \pi} \, \sum_{ k \ne 0 } \, 
    \sqrt{ \pi | k | } \,
    ( - \hat{a}_k         \, e^{   i k \sigma }
    + \hat{a}_k^\dagger \, e^{ - i k \sigma } )
    \label{3.12}
\end{eqnarray}
and the oscillators $ [ \, \hat{a}_k , \hat{a}_l^\dagger ] = \delta_{ k l } $
are represented on the Fock space $ {\cal F} $.

Combining the above parts,
the fundamental algebra without the central extension
is represented by the tensor product
$ {\cal H}_{s} \otimes {\cal H}_{t} \otimes {\cal F} $.
Thus it is concluded that
there are an infinite number of inequivalent quantizations
of the toroidal boson which are parametrized by $ 0 \le s, t < 1 $.

Now we reintroduce the radius of the target space $ R $.
Then we make replacement
$ \hat{X}( \sigma )  \to \frac{1}{R} \hat{X}( \sigma ) $ and
$ \hat{P}( \sigma )  \to          R  \hat{P}( \sigma ) $.
Accordingly (\ref{3.1}) and (\ref{3.2}) are replaced by
\begin{eqnarray}
	&&
	\hat{V} ( \sigma ) 
	= 
	e^{ i \frac1R \hat{N} \sigma } 
	\hat{V} 
	e^{ i \frac1R \hat{x} ( \sigma ) },
	\label{3.13}
	\\
	&&
	\hat{U}_f 
	= 
	\hat{W}^m e^{ -i R \lambda \hat{P} } 
	\exp 
	\left[ 
		-i R \int_0^{2 \pi} \tilde{f}( \sigma ) \hat{\pi}( \sigma ) d \sigma 
	\right],
	\label{3.14}
\end{eqnarray}
namely, the substitution
$ \hat{N} \to \frac1R \hat{N} $,
$ \hat{x}( \sigma ) \to \frac1R \hat{x}( \sigma ) $,
$ \hat{P} \to R \hat{P} $ and
$ \hat{\pi}( \sigma ) \to R \hat{\pi}( \sigma ) $
is made.
The eigenvalues of $ \hat{N} $ and $ \hat{P} $ are rescaled;
\begin{eqnarray}
	&&
    \hat{N} \, | n + s \ket 
    = 
    R ( n + s ) \, | n + s \ket,
    \label{3.15}
    \\
    &&
    \hat{P} \, | p + t \ket 
    = 
    \frac1R ( p + t ) \, | p + t \ket.
    \label{3.16}
\end{eqnarray}
In this case the Hamiltonian is given by
\begin{eqnarray}
	\hat{H}
    & = &
    \frac12 \int_0^{ 2 \pi }
    \left[
        \hat{P}^2 ( \sigma )
        +
        \Bigl( \frac{ d \hat{X} }{ d \sigma } \Bigr)^2
	\right]
    d \sigma
    \nonumber
    \\
    & = &
    \frac12 
    \Bigl( \frac{1}{ 2 \pi } \hat{P}^2 + 2 \pi \hat{N}^2 \Bigr)
    +
    \sum_{ k \ne 0 } | k | \hat{a}_k^\dagger \hat{a}_k,
    \label{3.17}
\end{eqnarray}
where we have substitute
$ \hat{X}( \sigma ) = \hat{N} \sigma + \hat{x}( \sigma ) $ and
$ \hat{P}( \sigma ) = \frac{1}{2 \pi} \hat{P} + \hat{\pi}( \sigma ) $.
So excitations of zero modes and winding modes exhibit the spectrum
\begin{equation}
	\hat{H} 
    | \, n + s ; \, p + t \ket
    = 
    \frac12 
    \left[
    	\frac{1}{ 2 \pi R^2 } (p+t)^2 
    	+ 2 \pi R^2 (n+s)^2 
    \right]
    | \, n + s ; \, p + t \ket.
    \label{3.18}
\end{equation}
Hence the spectrum is invariant under the transformation
$ R \to 1/( 2 \pi R ) $ if and only if $ s = t $ or $ -t $ (mod 1).
So the target space duality
which is expected from classical geometry~\cite{OAlvarez2}
becomes a rather subtle concept in quantum theory.

We would like to mention another restriction imposed on the values
of $ s $ and $ t $ from the viewpoint of string theory.
When we regard our model as a string theory,
the global diffeomorphisms require
the modular invariance of the partition function
$ Z ( \tau ) = \mbox{Tr} \exp ( 2 \pi i \tau \hat{H} ) $.
Then the modular invariance restricts $ s $ and $ t $
to be integers~\cite{Sakamoto}.
%
\section{Representations with $ d \ne 0 $}
When the nonvanishing central extension (\ref{2.17}) exists,
both the fundamental algebra and its representations are modified.
In our previous work~\cite{sigma} we considered only the case of $ d = 1 $.
Here we consider all possible values of $ d \in \Z $.

At first we change the factorization of $ U_f $ from (\ref{3.2}) to
\begin{equation}
	\hat{U}_f =  
	e^{ i d m \lambda } \, \hat{W}^m \, e^{ -i \lambda \hat{P} } 
	\exp 
	\left[ 
		-i \int_0^{2 \pi} \tilde{f}( \sigma ) \hat{\pi}( \sigma ) d \sigma 
	\right],
	\label{4.1}
\end{equation}
Then the modified algebra (\ref{2.13}) with the center (\ref{2.17}) is
satisfied if we add new commutation relations
\begin{eqnarray}
	&&
	[ \, \hat{P} , \hat{W} \, ] = - 2 d \hat{W},
	\label{4.2}
	\\
	&&
	[ \, \hat{\pi}( \sigma ) , \hat{\pi}( \sigma' ) \, ] 
	= \frac{ i d }{ \pi } \delta' ( \sigma - \sigma' ),
	\label{4.3}
\end{eqnarray}
to (\ref{3.4})--(\ref{3.6}).
The first one (\ref{4.2}) implies that 
the zero mode momentum $ \hat{P} $ is decreased by $ 2 d $ units 
when the winding number $ \hat{N} $ is increased by one unit
under the operation of $ \hat{W} $.
We may call this commutator a ``twist''.
Such an interrelation between the zero mode and winding number was unexpected
and its dynamical meaning is still obscure.
The second equation (\ref{4.3}) reminds us of the anomalous commutator
of the Schwinger model.

Representations of the above algebra are easy to construct but
considerably different from the previous ones.
The algebra (\ref{3.4}) and (\ref{3.5}) twisted with (\ref{4.2})
is represented by
\begin{eqnarray}
	&&
    \hat{N} \,  | \, n + s ; \, p + t \ket
    = ( n + s ) | \, n + s ; \, p + t \ket,
    \label{4.4}
    \\
    &&
    \hat{W} \, | \, n     + s ; \, p       + t \ket
    =          | \, n + 1 + s ; \, p - 2 d + t \ket,
    \label{4.5}
    \\
	&&
    \hat{P}     \, | \, n + s ; \, p + t \ket
    = ( p + t ) \, | \, n + s ; \, p + t \ket,
    \label{4.6}
    \\
    &&
    \hat{V} \, | \, n + s ; \, p     + t \ket
    =          | \, n + s ; \, p + 1 + t \ket.
    \label{4.7}
\end{eqnarray}
The Hilbert space spanned by
$ \{ | \, n + s ; \, p + t \ket | \, n, p \in \Z \} $
is denoted by $ {\cal T}_{s t}^{ (d) } $.

Taking the anomalous commutator (\ref{4.3}) into account,
the Fourier expansion of $ \hat{x}( \sigma ) $, $ \hat{\pi}( \sigma ) $
into oscillators is changed. After a tedious calculation we obtain
\begin{eqnarray}
	&&
    \hat{x} (\sigma)
    =
    \sum_{ k \ne 0 } \, 
    \frac{1}{ \sqrt{ 2 |d k| } } \,
    ( \hat{a}_k         \, e^{   i k \sigma }   
    + \hat{a}_k^\dagger \, e^{ - i k \sigma } ),
    \label{4.8}
    \\
    &&
    \hat{\pi} (\sigma)
    =
    \frac{i}{2 \pi}
    \left\{
    	\begin{array}{ll}
        	\sum_{ k = 1 }^{ \infty } & (d>0) \\
            \sum_{ k =-1 }^{-\infty } & (d<0)   
        \end{array}
    \right\}
     \sqrt{ 2 |d k| }
    ( - \hat{a}_k         \, e^{   i k \sigma }
      + \hat{a}_k^\dagger \, e^{ - i k \sigma } ).
    \label{4.9}
\end{eqnarray}
Details of this calculation are shown in~\cite{Dron}.
It should be noticed that
only positive $ k $ oscillators (right-moving boson) appear
in the expansion of $ \hat{\pi} $
when $ d > 0 $,
while only negative $ k $ oscillators (left-moving boson) appear
when $ d < 0 $.
However both positive and negative $ k $'s appear in $ \hat{x} $.
Let us denote the Fock space by $ {\cal F}^{ (d) } $
in which half of the modes are separated according to the sign of $ d $.

It is concluded that the algebra with the the nonvanishing central extension
is represented by $ {\cal T}_{st}^{ (d) } \otimes {\cal F}^{ (d) } $.
The first implication of the central extension is
the twisted interrelation between the winding number and the zero mode.
The second one is the lack of half of the oscillator modes in
$ \hat{\pi}(\sigma) $.
%
\section{Conclusion}
Here we shall give a brief summary of our method and results.
We introduced the zero mode and the winding number variables
to describe the topological property of the model
in (\ref{3.1}) and (\ref{3.2}).
We next defined the fundamental algebra (\ref{2.11})--(\ref{2.13}) and
decomposed them into (\ref{3.4})--(\ref{3.6}).
Then we constructed its representations and showed the existence of
inequivalent representations parametrized by $ 0 \le s, t < 1 $.
The spectrum is shifted by $ s $ and $ t $
and target space duality remains only when $ s = t $ or $ -t $ (mod 1),
as shown in (\ref{3.18}).
It was pointed that 
the algebra can be deformed by the central extension (\ref{2.17}).
The deformed algebra leads to the twisted interrelation between
the zero mode quantum number $ p $ and the topological number $ n $
in (\ref{4.2}) and (\ref{4.5})
and also to the lack of half of the oscillators in $ \hat{\pi}( \sigma ) $,
as observed in (\ref{4.9}).

Our scheme depends on the abelian nature of the model;
since our model is the $ U(1) $ sigma model,
it is possible to decompose its degrees of freedom
in the rather simple way of (\ref{3.1}) and (\ref{3.2}).
However it seems difficult to find suitable quantum variables
to describe a nonlinear sigma model
which has a general Riemannian manifold as a target space.
Although in the context of classical geometry 
a neat consideration of target space duality is given 
to general nonlinear sigma models
from the viewpoint of canonical transformations~\cite{EAlvarez,OAlvarez1},
the implication of target space duality in the context of quantum theory
should be investigated more carefully.
%
\section*{Acknowledgments}
The author would like to thank I. Tsutsui
for his continuous encouragement and useful advice.
He also thanks to A. Bordner for careful reading of the manuscript.
This work was supported by the Grant-in-Aid for Scientific Research from
the Ministry of Education, Science and Culture (No. 00074364).
%

%
\end{document}